%% file: slacpub7941.tex
\documentclass{article}
\usepackage{ltwol2e}

\def\Journal#1#2#3#4{{#1} {\bf #2}, #3 (#4)}

\def\PLB{{\em Phys. Lett.}  B}
\def\PRL{\em Phys. Rev. Lett.}
\def\PRD{{\em Phys. Rev.} D}
\def\ZPC{{\em Z. Phys.} C}
\def\EPJ{{\em Eur. Phys. J.} C}
\def\CPC{\em Comp. Phys. Comm.}

\def\be{\begin{equation}}
\def\ee{\end{equation}}
\def\bea{\begin{eqnarray}}
\def\eea{\end{eqnarray}}

\input psfig

\begin{document}
\onecolumn
\pagestyle{empty}

\rightline{SLAC-PUB-7941}
\rightline{October 10, 1998}
\begin{center}
\vspace*{1.8cm}
{\large\bf Charmless and Double-Charm \mbox{\boldmath $B$} Decays at 
\renewcommand{\thefootnote}{\fnsymbol{footnote}}SLD\footnote{Work 
supported by U.S. Department of Energy contract DE-AC03-76SF00515.}}\\
\vspace*{6.0ex}
{\large Mourad Daoudi}\\
\vspace*{1.5ex}
{\it Stanford Linear Accelerator Center\\
Stanford, CA 94309}\\
\vspace*{0.4in}
Representing the SLD Collaboration$^{\ast\ast}$ \\
\vspace*{1.2in}
{\bf Abstract}\\
\ \\
\end{center}
\noindent
We present new results from the study of $B$ decays at SLD: a measurement 
of the inclusive double-charm branching fraction in $B$-hadron decays and 
a search for the rare exclusive charmless decays
$B^+\rightarrow\rho^0\pi^+(K^+)$ and $B^+\rightarrow K^{\ast 0}\pi^+(K^+)$. 
Using a novel technique which consists of counting charged kaons produced 
at the secondary ($B$) vertex and the tertiary ($D$) vertex, we measure 
${\cal B}(B\rightarrow D\bar DX) = 0.188\pm 0.025 (stat)\pm 0.059 (syst)$.
Another technique, based on the comparison of the charge of the $D$ vertex 
to the flavor of the $B$ hadron at production, yields a consistent result. 
In the search for rare exclusive $B^+$ decays, no candidates were 
found and competitive branching ratio upper limits are derived.

\vspace*{2.2in}
\begin{center}
{\small
Talk given at the XXIX International Conference on High Energy Physics\\
Vancouver, BC, Canada, July 23--29, 1998}
\end{center}
 
\newpage
\pagestyle{plain}


\title{CHARMLESS AND DOUBLE-CHARM B DECAYS AT SLD}
\author{M. DAOUDI}
\address{Stanford Linear Accelerator Center, Stanford, CA 94309, USA\\
E-mail: daoudi@slac.stanford.edu}

\twocolumn[\maketitle\abstracts{ 
We present new results from the study of $B$ decays at SLD: a measurement 
of the inclusive double-charm branching fraction in $B$-hadron decays and 
a search for the rare exclusive charmless decays
$B^+\rightarrow\rho^0\pi^+(K^+)$ and $B^+\rightarrow K^{\ast 0}\pi^+(K^+)$. 
Using a novel technique which consists of counting charged kaons produced 
at the secondary ($B$) vertex and the tertiary ($D$) vertex, we measure 
${\cal B}(B\rightarrow D\bar DX) = 0.188\pm 0.025 (stat)\pm 0.059 (syst)$.
Another technique, based on the comparison of the charge of the $D$ vertex 
to the flavor of the $B$ hadron at production, yields a consistent result. 
In the search for rare exclusive $B^+$ decays, no candidates were 
found and competitive branching ratio upper limits are derived.}]
\vspace{2.2in}
\section{Introduction}
The motivation for the analyses presented here lies in trying to shed some 
light into a long-standing $B$ decay puzzle. For some time, the
$B$ semileptonic branching ratio ($B_{SL}$) and the charm yield in $B$ decays
($n_C$) appeared to be too low compared to the theoretical 
expectation. While this is no longer the case at LEP,
this discrepancy seems to persist at the $\Upsilon$(4S), as 
illustrated in Fig~1~\cite{neubert,feindt}.
\par
One way of reconciling a small $B_{SL}$ is by having a large $B$ hadronic
width $\Gamma_{Had}$, which in turn can be due to a large 
$b\rightarrow c\bar cs$ partial
width. By the same token, a large $B\rightarrow D\bar DX$ branching ratio
can result in a large charm yield. This is shown in the following equations:
\begin{eqnarray*}
        B_{SL} & = & {\Gamma_{SL}\over{\Gamma_{SL} + \Gamma_{Had}}} \\
  \Gamma_{Had} & = & \Gamma(b\rightarrow c\bar ud) + 
                     \Gamma(b\rightarrow c\bar cs)   \\
               &   & \mbox{\hskip 0.68in} +
                     \Gamma(b\rightarrow sg) + \Gamma(b\rightarrow u) \\
           n_C & = & 1 + f_{D\bar D} - f_{NOC}.  
\end{eqnarray*}
Here, $f_{D\bar D}$ and $f_{NOC}$ represent the $B$ double-charm and 
no-open-charm branching fractions, respectively.
\begin{figure}[h]
\begin{center}
\hbox{\hskip 0.3cm \psfig{figure=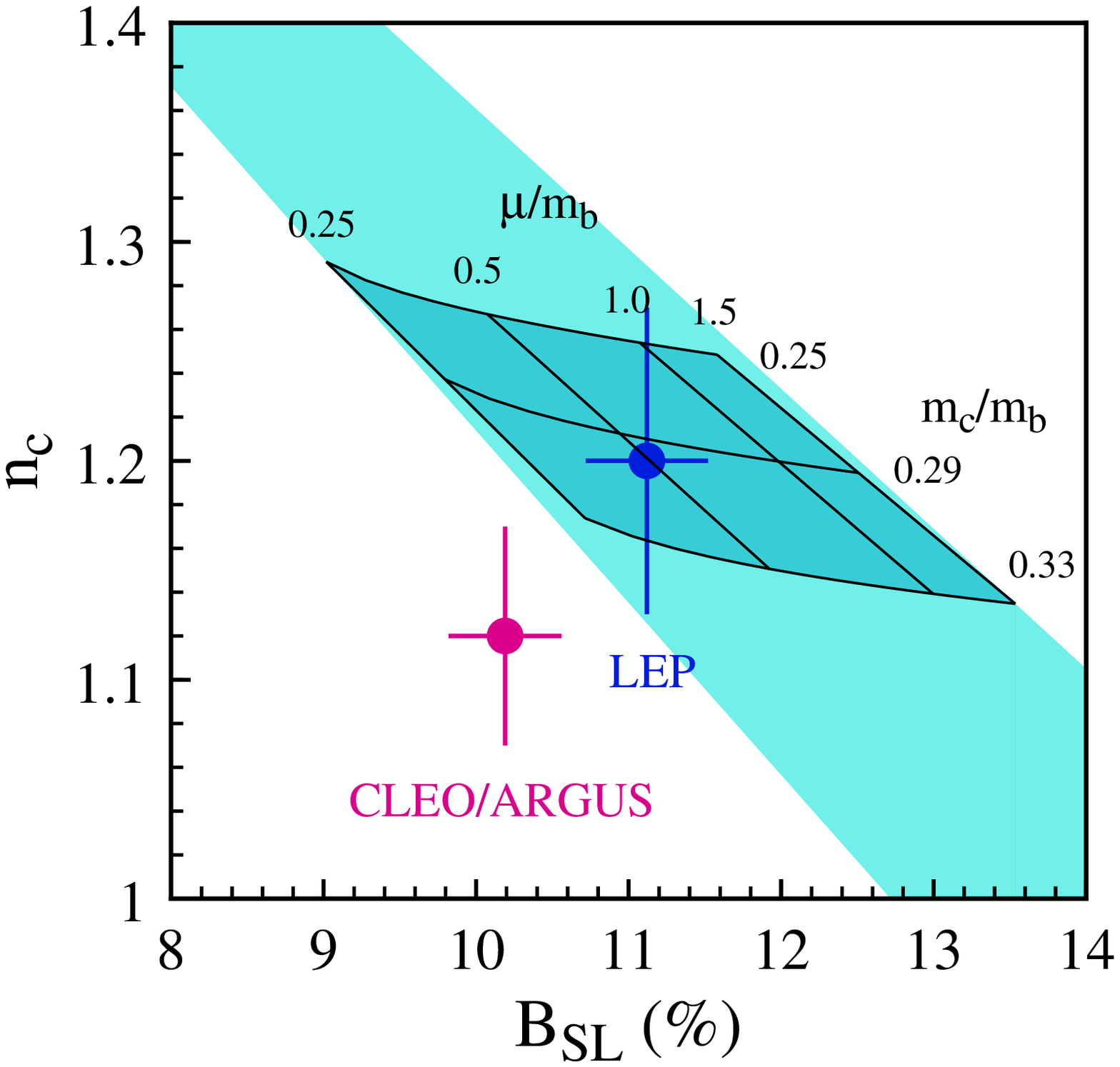,width=3.0in}}
\end{center}
\vspace{-0.4cm}
\centerline{\parbox{3.0in}{\footnotesize Figure 1: Charm yield vs. 
semileptonic branching ratio, from Ref.~1. Theoretical bounds are defined 
by the quark mass ratio $m_c/m_b$ and the renormalization scale $\mu$.}}
\end{figure}
\par
Another possible scenario is to have a large contribution from the 
$b\rightarrow sg$ and $b\rightarrow u$ transitions. For example, A. Kagan
and J. Rathsman suggest that the $b\rightarrow sg$ decay could possibly be
enhanced to a branching ratio as large as 10\%~\cite{kagan}. An inclusive 
search for $b\rightarrow sg$ is currently underway at SLD~\cite{sldbsg}. 
Preliminary results from that analysis were covered in A. Litke's 
talk~\cite{litke} at this conference.
\par
In the present paper, results are given on the measurement of the inclusive 
branching fraction ${\cal B}(B\rightarrow D\bar DX)$ and a search for the
rare exclusive decays $B^+\rightarrow\rho^0\pi^+(K^+)$ and 
$B^+\rightarrow K^{\ast 0}\pi^+(K^+)$. These analyses benefit from the very
good tracking and vertexing efficiency and resolution of SLD provided by a
precision 3-D CCD pixel vertex detector. In addition, the high polarization of
the SLC electron beam permits an excellent $b / \bar b$ separation, while the
Cherenkov Ring Imaging Detector (CRID) provides good particle 
identification. The total SLD data sample of $\sim 550k$
hadronic $Z^0$'s, collected between 1993 and 1998, was used in the rare 
$B^+$ decay search, whereas $\sim 250k$ hadronic $Z^0$'s were analyzed in the
$B$ double-charm fraction measurement.
\section{Measurement of \boldmath{${\cal B}(B\rightarrow D\bar DX)$}}
The $B\rightarrow D\bar DX$ decay channel is governed mainly by the 
$b\rightarrow c\bar cs$ quark transition illustrated by the spectator
diagram of Fig.~2, where in addition to the $D$ meson from the $b\rightarrow c$
transition, a second so-called ``wrong-sign'' $D$ is produced in the
$W^-\rightarrow\bar cs$ decay.
\begin{table*}[t]
\caption{Number of charged kaons in $B$- and $D$-vertex and their ratio.
\label{tab:kaon}}
\vspace{0.2cm}
\begin{center}
\begin{tabular}{lcrrcrrcl} \hline\hline
  && $N_{K^-,B}$   & $N_{K^+,B}$  && $N_{K^-,D}$  & $N_{K^+,D}$ && 
\hbox{\hskip 0.1cm}  $N_{K,B} / N_{K,D}$ \\ \\ \hline
  Data     &&  758 &  559    && 1168 &  746   &&  $0.688\pm 0.025$ \\ \\
  M.C. $D\bar D$  && 3115 & 2339 && 2380 &  1880  && $1.280\pm 0.026$ \\ \\
  M.C. ``Not-$D\bar D$'' && 5386 & 3974 && 11489 & 6549 && $0.519\pm 0.007$ \\ 
\hline\hline
\end{tabular}
\end{center}
\end{table*}
Previous measurements~\cite{cleobdd,alephbdd,delphibdd} of the branching
ratio for this decay yielded a value of $\sim 20\%$ with an uncertainty 
$\simeq\pm 4\%$. 
\par
At SLD, the measurement is performed using an inclusive approach based on 
selecting $B$ decays with
two well separated vertices. This is achieved, in part, by requiring that
the $\chi^2$ probability that all tracks in the $B$ decay chain originate
from a single vertex be $< 5\%$. We refer to the vertex closest to the
interaction point as the ``$B$-vertex'' and the one furthest as the 
``$D$-vertex'', as in the standard $b\rightarrow c$ decays. Note that for 
$B\rightarrow D\bar DX$ decays, the so-called ``$B$-vertex'' is rather 
likely to be one of the two $D$ mesons. The two unique techniques we make 
use of to extract the double-charm fraction $f_{D\bar D}$ rely essentially 
on this fact. They are described below.\\
\psfig{figure=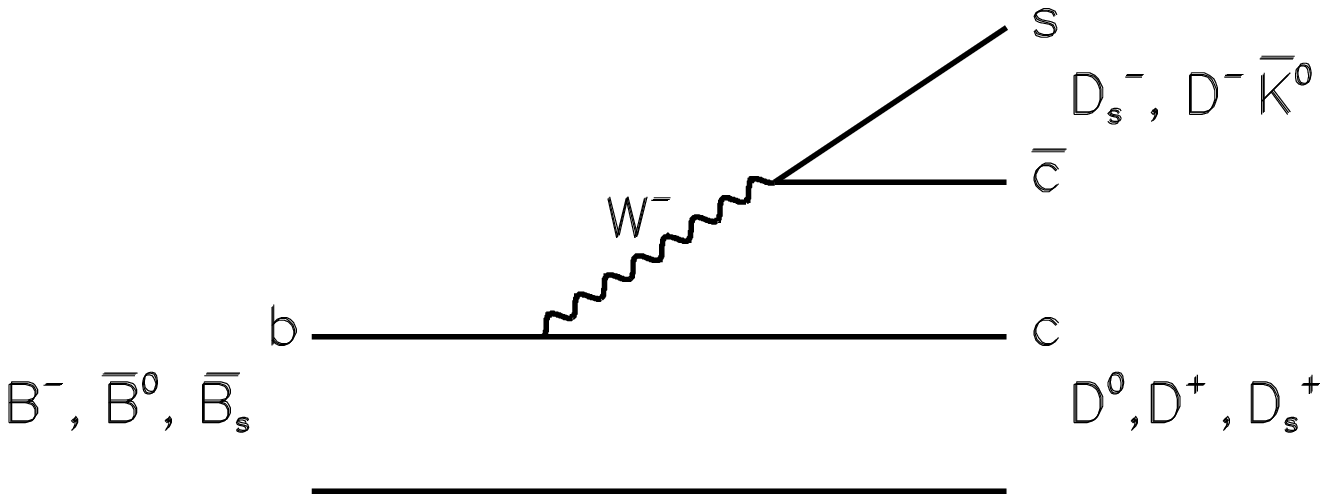,width=3.5in} \ \\
\vskip -0.9cm
\centerline{{\footnotesize Figure 2: Tree-level diagram for 
$b\rightarrow c\bar cs$.}}
\begin{figure}[ht]
\begin{center}
\psfig{figure=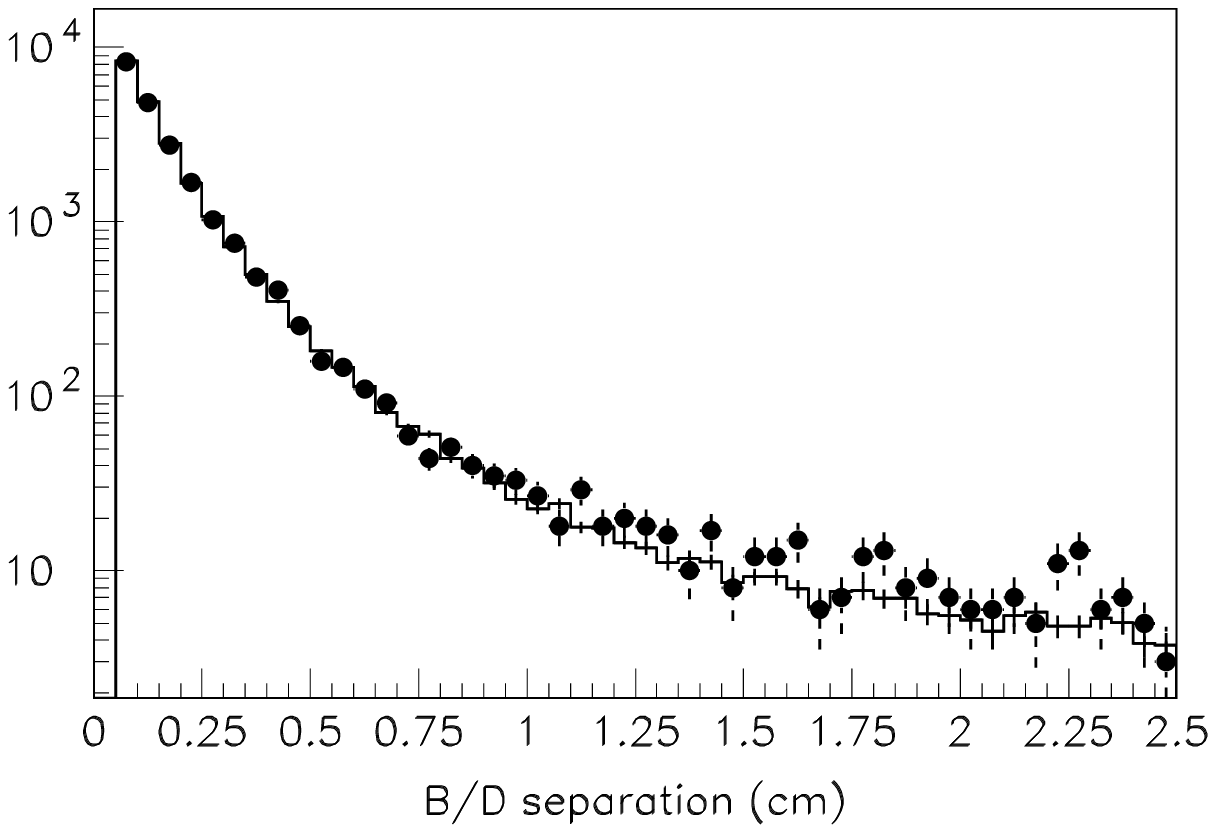,width=3.4in}
\parbox{3.2in}{\footnotesize Figure 3: Reconstructed distance between $B$ and 
$D$ vertices. The dots represent the data and the histogram the Monte Carlo.}
\end{center}
\end{figure}
\par 
Fig.~3 shows the reconstructed distance between the $B$ and $D$
vertices. Good agreement between the data and the Monte Carlo is seen. 
The Monte Carlo simulation of the production and decay of $B$ mesons at
the $Z^0$ resonance is based on the JETSET 7.4
generator~\cite{jetset} and the CLEO $B$ decay model~\cite{bmodel}.
\subsection{The Kaon Counting Method}
In this analysis we count the number of charged kaons identified by the CRID
that are attached to the
$B$-vertex and the $D$-vertex. We then compare the ratio of these two numbers
in the data to that from two Monte Carlo samples: one containing exclusively
double-charm $B$ decays, and one made up of all other $B$ decay channels. 
Since kaons are copiously produced in the $c\rightarrow s$ transition,
one expects to find substantially more kaons in the $D$-vertex for single-charm
$B$ decays. This is not the case for double-charm $B$ decays where
one of the $D$ mesons may be reconstructed as the
$B$-vertex. The ratio of the number of $K^\pm$'s in the $B$-vertex to that in
the $D$-vertex is shown in the last column of Table~\ref{tab:kaon}. It can
be seen clearly that this ratio in the data has a double charm component, 
whose fraction we deduce to be
\[ f_{D\bar D} = 0.188\pm 0.025 (stat). \]
This is after having applied a correction for the difference in 
efficiency for
selecting single-charm ($\epsilon_D$) and double-charm ($\epsilon_{D\bar D}$)
decays of the $B$. We estimate the ratio of these efficiencies to be
$\epsilon_{D\bar D} / \epsilon_D = 1.34\pm 0.03$.
\par 
An interesting cross check is provided by the $K^+/K^-$ asymmetry at
the $D$-vertex (see Table~\ref{tab:kaon}). For a single-charm $B$ decay, 
the sign of the kaon there
is -- (+) if it originated from the $b\rightarrow c\rightarrow s$
($\bar b\rightarrow\bar c\rightarrow\bar s$) decay chain. For double-charm 
decays, the $D$-vertex may correspond to the wrong-sign $D$ meson which
produces a kaon with opposite sign. Thus the kaon charge asymmetry
in this case is different.
The $B\rightarrow D\bar DX$ fraction we extract from this asymmetry
is $f_{D\bar D} = 0.29\pm 0.11(stat)$.
\par 
Note that, in this discussion, the $K^\pm$ sign is given by 
the kaon charge multiplied by the sign of the $b$-quark charge,
as determined by tagging the $b$ or $\bar b$ flavor of the $B$ meson at 
production. The initial state $b$ flavor tagging at SLD relies mainly on the 
large polarized forward-backward asymmetry for $Z^0\rightarrow b\bar b$ decays.
It is complemented by other tags using information from the opposite hemisphere
to the selected $B$ vertex. These tags include the momentum-weighted track 
charge sum (jet charge), the dipole charge for a topologically reconstructed
vertex and the charge of an identified kaon or a high transverse-momentum
lepton. Further details can be found in Ref.~11.
\par
The analyzing power of the kaon counting technique is somewhat reduced 
experimentally due to the misassignment of tracks between the $B$ and 
$D$ vertices. In order not to rely entirely on the Monte Carlo simulation, 
the track misassignment rate is calibrated in the data using semileptonic 
$b\rightarrow l$ and $b\rightarrow c\rightarrow l$ decays. This is done by 
comparing the lepton
sign in these decays to the initial-state $b$-quark flavor (sign). Wrong 
combinations of the two are the result of an initial-state flavor mistag 
(whose rate is well determined) and/or a misassignment of the lepton to the 
$B$- or $D$-vertex. We measure the track assignment 
efficiency to be $\epsilon_{T.A.} = 0.70\pm 0.03$. The same analysis performed
on the Monte Carlo gives a consistent result. The statistical error in this
efficiency translates into a systematic uncertainty of $\pm 0.042$ 
in $f_{D\bar D}$. 
\begin{table}[t]
\begin{center}
\parbox{3.0in}{\footnotesize Table 2: Systematic errors in the measurement of 
$f_{D\bar D}$ for the kaon counting method.\\
\ }
\vspace{0.2cm}
\begin{tabular}{lc} \hline\hline 
&  $\sigma_{\rm syst}$  \\  Detector Systematics  &  \\ \hline
Tracking Efficiency & $\pm 0.003$ \\
Tracking Resolution & $\pm 0.007$ \\ 
$K^\pm$ Id. Efficiency & $< 0.001$ \\
$\pi^\pm$ Fake Rate &   $\pm 0.007$ \\
Track Misassignment \hbox{\hskip 0.7cm} & 
 $\pm 0.042$  \\  \\
 Physics Systematics &  \\ \hline
$udsc$ Background & $\pm 0.004$ \\
$B$ Multiplicity  &  $\pm 0.015$  \\
$B_s$ Fraction    & $\pm 0.011$ \\
$\Lambda_b$ Fraction    & $\pm 0.007$ \\
${\cal B}(B_{u,d}\rightarrow D^0X)$ & $\pm 0.002$ \\
${\cal B}(B_{u,d}\rightarrow D^+X)$ & $\pm 0.002$ \\
${\cal B}(B_{u,d}\rightarrow D^+_sX)$ & $\pm 0.004$ \\
$B\rightarrow D\bar D$ Model & $\pm 0.015$  \\
Charm Multiplicity &  $\pm 0.032$  \\ \\
\hbox{\hskip 0.7cm} TOTAL &  $\pm 0.059$ \\ \hline\hline
\end{tabular}
\end{center}
\end{table}
\par
Another important source of systematic error in this measurement is the
uncertainty in the charm decay multiplicity, which is not very well determined
experimentally~\cite{markthree}. Other systematic errors are listed in 
Table~2. Combining all the components in quadrature we get
a total systematic error of $\pm 0.059$. Thus, our preliminary measurement
of the $B$ double-charm branching fraction using the kaon counting method is
\[ f_{D\bar D} = 0.188\pm 0.025 (stat)\pm 0.059(syst). \]
\subsection{$D$ Charge Method} 
In this analysis, we compare the sign of the reconstructed $D$-vertex to the 
initial-state flavor of the $B$ meson. For example, when a $B^-$ decays to 
a single charmed meson, we expect the $D$-vertex to be positively charged
(see Fig.~2). 
Whereas, for a double-charm decay of the same $B^-$, the reconstructed 
$D$-vertex can be either positively or negatively charged, since it may
correspond to either a right- or wrong-sign $D$ meson. This is, of course,
without taking into account decays of the $B^-$ into a neutral $D$ meson,
for simplicity. The distribution of
the $D$-vertex charge multiplied by the $b$-quark charge sign of the 
$B$ meson at production is shown in Fig.~4. There is good agreement
between the data, represented by the dots, and the Monte Carlo, by the solid 
histogram. The Monte Carlo is a sum of a pure single-charm (shaded histogram) 
component and double-charm component. Note that the latter is asymmetric.
\vskip 0.2cm
\begin{center}
\psfig{figure=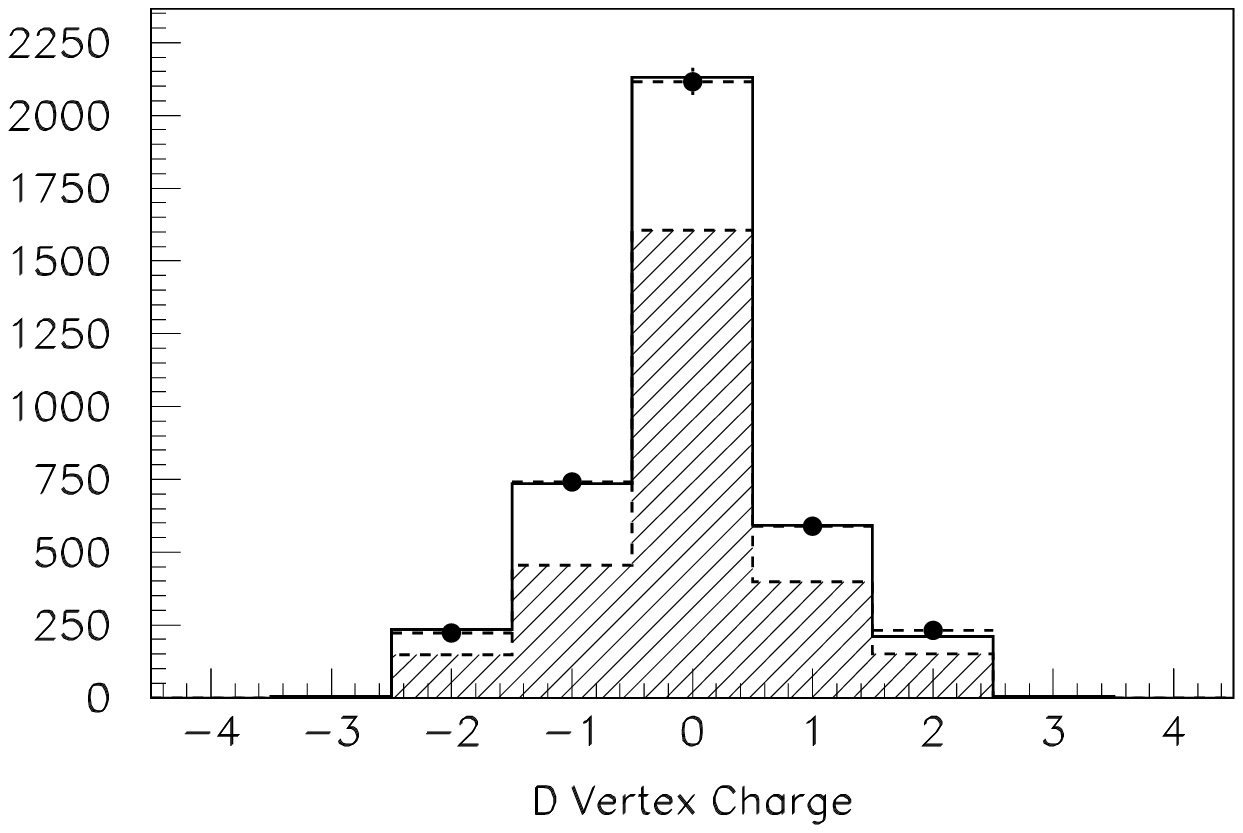,width=3.5in}
\parbox{3.2in}{\footnotesize Figure 4: $D$-vertex charge (multiplied by 
sign of $b$-quark charge). The various components are explained in the text.}
\vskip 0.2cm
\end{center}
\par
Similarly to the kaon counting method, this technique is also sensitive to
the misassignment of tracks between the $B$ and $D$ vertices. Furthermore,
the large $D^0$ production rate degrades its analyzing power.
\par
We extract the $B$ double-charm branching fraction by fitting the data
distribution in Fig.~4 to the two Monte Carlo distributions
corresponding to the $D$-charge for single- and double-charm decays. The result
of the $\chi^2$ fit is
\[ f_{D\bar D} = 0.195\pm 0.045 (stat), \]
which is consistent with the above value obtained with the kaon counting 
method. The systematic error for this measurement is currently being evaluated.
\section{Search for Charmless \boldmath{$B^+$} Decays}
We have searched for the exclusive three-prong decays
 $B^+ \rightarrow\rho^0\pi^+(K^+)$,  
 $\rho^0\rightarrow\pi^+\pi^-$; and
 $B^+ \rightarrow K^{\ast 0}\pi^+(K^+)$, 
 $K^{\ast 0}\rightarrow K^+\pi^-$.
These are mediated by the tree-level $b\rightarrow u$ and the one-loop penguin
$b\rightarrow d(s)$ transitions. Theoretical predictions for the corresponding
branching ratios are in the $10^{-6}$ -- $10^{-5}$ 
range~\cite{deshpande,chau,deandrea}. The best experimental
limits come from CLEO~\cite{cleobu} and are about a factor 5 -- 10 larger.
These are based on the observation of a small number of events (e.g. 8 
observed decays for an expected background of 3 in the $\rho\pi$ channel).
For the decays $B^+\rightarrow\rho^0\pi^+, \ K^{\ast 0}\pi^+$,
the DELPHI~\cite{delphibu} experiment observes a total of 3 decays, with a
background estimated at 0.15. The resulting branching ratio is
${\cal B}(B\rightarrow\rho (K^\ast)\pi) = (1.7^{+1.2}_{-0.8}\pm 0.2)\times
10^{-4}$.
\begin{figure}[h]
\begin{center}
\hbox{\hskip 0.4cm \psfig{figure=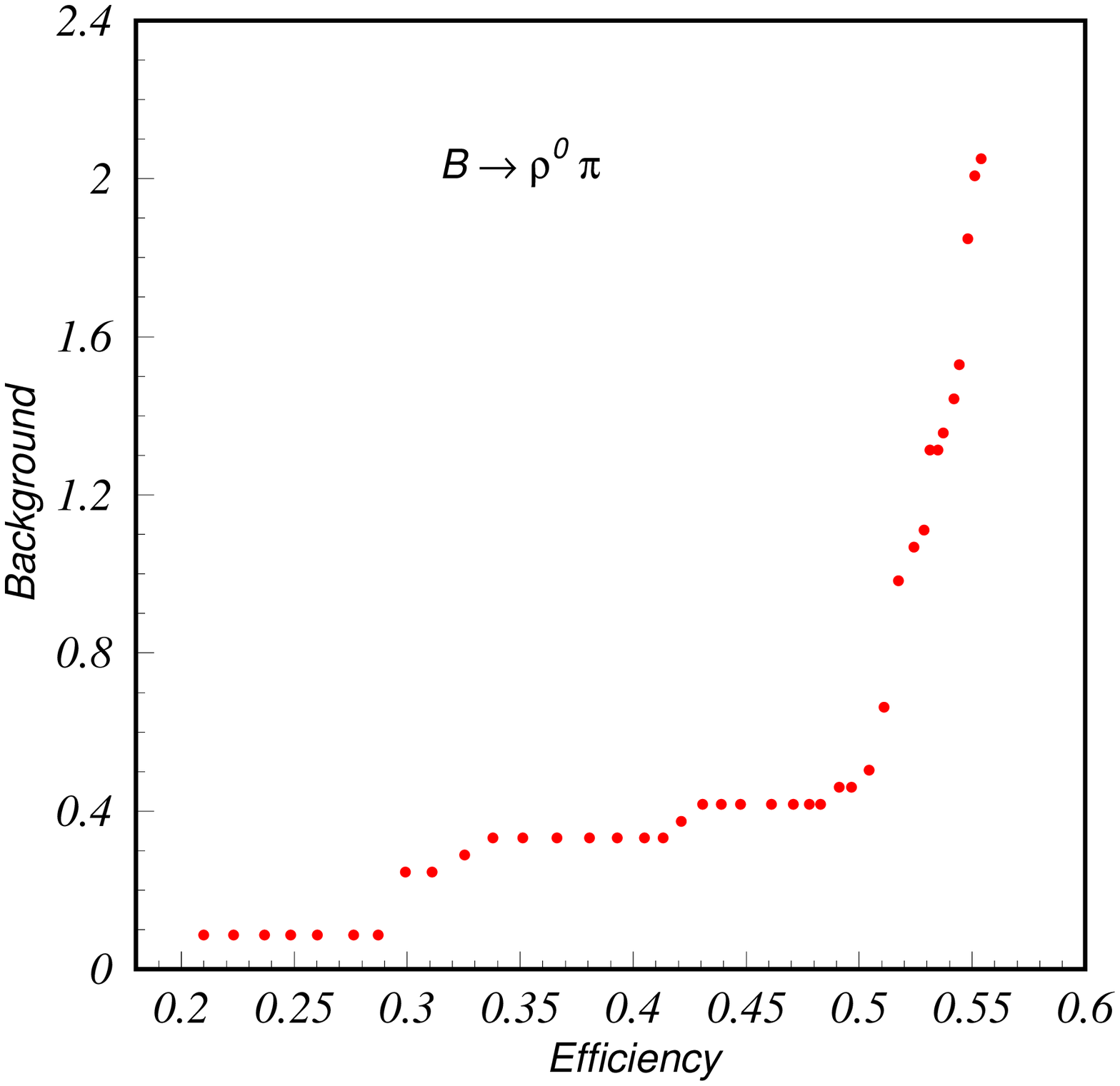,width=3.0in,height=2.2in}}
\vspace{0.2cm}
\parbox{3.0in}{\footnotesize Figure 5: Background vs. efficiency evolution as
the cut on the discriminator function output is varied.}
\end{center}
\end{figure}
\par
While CLEO and DELPHI have used relatively large samples of $B^+$'s, their
efficiency is small ($\approx 6.0\%$). At SLD, with our high resolution
vertexing capability, we achieve
an efficiency as high as 50\% within our acceptance.
With a sample of $\sim 100k$ $B^+$'s, we can set comparable limits to
CLEO and DELPHI.
\begin{table}[h]
\begin{center}
\parbox{3.0in}{\footnotesize Table 3: Efficiency and background estimate for 
each of the four final states in the $B^+\rightarrow$ 3-prong search
(not including the branching ratio for $\rho^0\rightarrow\pi^+\pi^-$ 
and $K^{\ast 0}\rightarrow K^+\pi^-$).}
\ \vskip 0.3cm
\begin{tabular}{lcccc}
                  && Efficiency  && Estimated  \\
                  &&             && Background \\ \hline
  $\rho^0\pi^+$   &&  0.50  && 0.61    \\
  $\rho^0 K^+$    &&  0.45  && 0.61    \\
  $K^{\ast 0}\pi^+$ && 0.52 && 0.37    \\
  $K^{\ast 0} K^+$  && 0.46 && 0.37    
\end{tabular}
\end{center}
\end{table}
\begin{figure}[h]
\begin{center}
\hbox {\hskip 1.0cm \psfig{figure=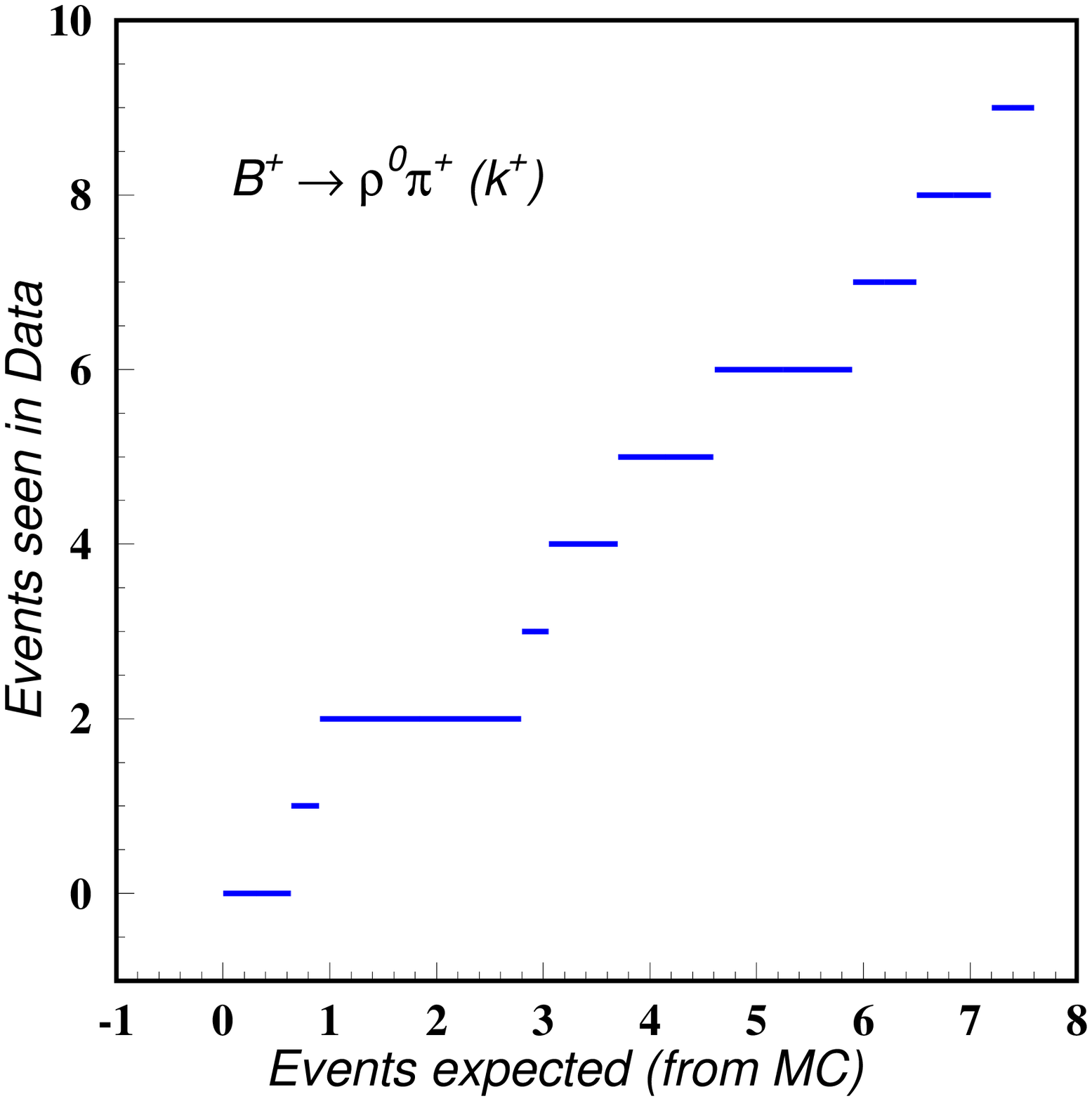,width=2.5in,angle=-90}}
\parbox{3.0in}{\footnotesize Figure 6: Number of observed 
decays in the data vs. expected background from the Monte Carlo, as the cut
of the discriminator function is relaxed.}
\end{center}
\end{figure}
\par
The analysis consists of three steps. First, we form a three-prong vertex
with charge $=\pm 1$. We apply track quality, background suppression, and 
decay kinematics requirements. The latter include $B$ mass and energy cuts, 
and $\rho^0$ and $K^{\ast 0}$ mass and center-of-mass helicity angle cuts.
No kaon identification is required. 
In the second stage of the analysis, we construct a discriminator function 
using eight kinematic and vertexing variables, each with a parametrization for 
signal and background. 
Finally, we determine the optimal cut on the output variable of this 
discriminator function. This is done, relying on Poisson statistics, 
by minimizing the expected branching ratio upper limit as a function 
of efficiency, for each of the decay channels, under the assumption
that the true branching fraction is 0. The variation of the expected
background vs. efficiency for the $\rho\pi$ channel is shown in Fig.~5.
The optimal discriminator function output cut for this channel can be 
inferred from the figure, corresponding to an efficiency of $\approx 50\%$.
It is important to note that this study is performed using only a Monte Carlo 
sample for the background, independently of the data.
\begin{table}[h]
\small
\begin{center}
\parbox{3.0in}{\footnotesize Table 4: Systematic uncertainty in the efficiency 
for each of the $B^+\rightarrow$ 3-prong final states.\\ 
\ }
\vspace{0.2cm}
\begin{tabular}{l|llll}
& $\rho^0\pi^+$ & $\rho^0 K^+$ & $K^{\ast 0}\pi^+$ & $K^{\ast 0} K^+$ \\ \hline
Mass Resolution      &  -0.3\% &  -0.9\% &  -0.4\% &  +0.1\% \\
Tracking Effic.  &  -0.6 &  -0.5 &  -0.4 &  -0.2 \\
Vertex  Resolution   &  -2.5 &  -1.7 &  -1.0 &  -0.9 \\
$b$ Fragmentation & $\pm 0.4$ & $\pm 0.0$ & $\pm 0.3$ & $\pm 0.4$\\ \hline
Total Uncertainty & $\pm 3.8\%$ & $\pm 3.1\%$ & $\pm 2.1\%$ & $\pm 1.4\%$  
\end{tabular}
\end{center}
\end{table}
\normalsize
\par
No events are observed in any of the four decays we searched for in the data.
The amount of background we estimate for each channel is listed in Table~3,
together with the individual efficiencies. 
The number of observed decays in the data is compared to the expected 
background as the cut on the discriminator function is relaxed.
We find that the two track one another very well, providing a nice cross
check of the analysis. This is illustrated in Fig.~6.
\par
Table~4 contains the systematic uncertainty in the efficiency
for each of the decay channels. The first three components in the table 
correspond to the shift in absolute efficiency observed in the Monte Carlo
when it is smeared according to transverse momentum resolution, tracking
efficiency, and impact parameter resolution, respectively. The last component
corresponds to the uncertainty associated with the measurement of the average
$B$-hadron energy at the $Z^0$.
Taking into account these systematic errors and the uncertainty in the
estimate of the number of $B^+$'s contained in the data ($\pm 1.1\%$),
we set 90\% C.L. upper limits. These are given in Table~5. They are 
consistent with the CLEO limits. For the combined $\rho^0\pi^+$ and 
$K^{\ast 0}\pi^+$ final states, the agreement between our result and DELPHI's 
branching ratio measurement, given above, is only at the $\leq 10\%$ 
confidence level.
\vspace{0.1cm}
\begin{center}
\begin{tabular}{lccc} 
\multicolumn{4}{c} {\parbox{3.0in}
{\footnotesize Table 5: Preliminary 90\% CL upper limits 
(B.R.$\times 10^5$) for 
the four exclusive final states in the $B^+\rightarrow$ 3-prong search. 
Corresponding CLEO and DELPHI limits are also given.\\ 
\ } }
\\ \hline\hline
   & CLEO & DELPHI & SLD \\
   &      &        & Preliminary \\ \hline
   $B^+\rightarrow\rho^0\pi^+$     & 5.8 & 16 & \mbox{\hskip 0.15cm}8.2 \\
   $B^+\rightarrow\rho^0 K^+$      & 1.4 & 12 & \mbox{\hskip 0.15cm}9.2 \\
   $B^+\rightarrow K^{\ast 0}\pi^+$ & 3.9 & 39 & 11.9 \\
   $B^+\rightarrow K^{\ast0} K^+$   &     &    & 13.5 \\ 
\hline\hline
\end{tabular}
\end{center}
\section{Conclusions}
We have presented a measurement of the $B\rightarrow D\bar DX$ branching ratio 
using two novel techniques. These rely on the high efficiency inclusive 
topological vertex reconstruction of the SLD detector, as well as its 
excellent initial-state $b$ / $\bar b$ flavor tagging, and its very good 
particle identification capability. 
The first method, called the kaon counting technique, gives the following 
preliminary result:
\[ {\cal B}(B\rightarrow D\bar DX) = 0.188\pm 0.025 (stat)\pm 0.059(syst). \]
This is comparable in precision and in good agreement with recent measurements 
by other experiments. The systematic error, which is conservatively estimated 
at this stage of the analysis, is dominated by the track misassignment rate 
between secondary and tertiary vertices, and by the charm decay multiplicity. 
It is expected to be improved significantly.
\par 
The second method, dubbed the $D$-charge technique, gives a consistent 
result. This analysis is not as mature yet, but it provides a good cross
check for the kaon counting technique. The two measurements will
be combined, taking into account common systematics, and the overall error
in the double-charm branching ratio will be significantly decreased. An
additional improvement in the statistical error is expected with the inclusion
of a data sample not yet analyzed.
\par
We have also shown the results of a search for rare exclusive charmless
$B^+\rightarrow$ 3-prong decays. Here also, the high resolution tracking 
and vertexing of SLD greatly enhances the experimental sensitivity to these 
modes. An overall efficiency of $50\%$ is achieved. No candidates were 
observed in the data for the $\rho^0\pi^+(K^+)$ and $K^{\ast0}\pi^+(K^+)$ 
final states, and competitive upper limits are derived. 
\par
Other interesting analyses in $B$ decays are underway at SLD, in particular
a measurement of the $B$ semileptonic branching fraction and a search for 
the $b\rightarrow sg$ transition. Together with the measurements presented 
here, a significant contribution toward a better understanding
of the semileptonic branching ratio vs. charm yield puzzle will be made.
\section*{Acknowledgements}
I would like to thank my colleagues Glen Crawford, Per Reinertsen, and Bruce
Schumm for their contribution.

\newpage
\onecolumn
\include{sldauth97}
\end{document}

%% file: sldauth97.tex
\centerline{{\large\bf $^{\ast\ast}$ The SLD Collaboration}}
\vskip 0.5cm
\begin{center}
  \def\iADEL{$^{(1)}$}
  \def\iBOL{$^{(2)}$}
  \def\iBU{$^{(3)}$}
  \def\iBRUN{$^{(4)}$}
  \def\iUCSB{$^{(5)}$}
  \def\iUCSC{$^{(6)}$}
  \def\iCIN{$^{(7)}$}
  \def\iCSU{$^{(8)}$}
  \def\iCOLO{$^{(9)}$}
  \def\iCOL{$^{(10)}$}
  \def\iFER{$^{(11)}$}
  \def\iFRA{$^{(12)}$}
  \def\iILL{$^{(13)}$}
  \def\iLBL{$^{(14)}$}
  \def\iMIT{$^{(15)}$}
  \def\iMASS{$^{(16)}$}
  \def\iMISS{$^{(17)}$}
  \def\iNAG{$^{(18)}$}
  \def\iOREG{$^{(19)}$}
  \def\iPAD{$^{(20)}$}
  \def\iPERU{$^{(21)}$}
  \def\iPISA{$^{(22)}$}
  \def\iRUT{$^{(23)}$}
  \def\iRAL{$^{(24)}$}
  \def\iSOGANG{$^{(25)}$}
  \def\iSOONG{$^{(26)}$}
  \def\iSLAC{$^{(27)}$}
  \def\iTENN{$^{(28)}$}
  \def\iTOH{$^{(29)}$}
  \def\iVAND{$^{(30)}$}
  \def\iWASH{$^{(31)}$}
  \def\iWISC{$^{(32)}$}
  \def\iYALE{$^{(33)}$}
  \def\dead{$^{\dag}$}
  \def\andgen{$^{(a)}$}
  \def\andper{$^{(b)}$}
%
%
\mbox{K. Abe                 \unskip,\iNAG}
\mbox{K. Abe                 \unskip,\iTOH}
\mbox{T. Abe                 \unskip,\iSLAC}
\mbox{T. Akagi               \unskip,\iSLAC}
\mbox{N.J. Allen             \unskip,\iBRUN}
\mbox{D. Aston               \unskip,\iSLAC}
\mbox{K.G. Baird             \unskip,\iMASS}
\mbox{C. Baltay              \unskip,\iYALE}
\mbox{H.R. Band              \unskip,\iWISC}
\mbox{T. Barklow             \unskip,\iSLAC
% ...
}
\it
  \vskip \baselineskip                   
  \vskip \baselineskip                   
  \iADEL
     Adelphi University,
     Garden City, New York 11530 \break
  \iBOL
     INFN Sezione di Bologna,
     I-40126 Bologna, Italy \break
  \iBU
     Boston University,
     Boston, Massachusetts 02215 \break
  \iBRUN
     Brunel University,
     Uxbridge, Middlesex UB8 3PH, United Kingdom \break
  \iUCSB
     University of California at Santa Barbara,
     Santa Barbara, California 93106 \break
  \iUCSC
     University of California at Santa Cruz,
     Santa Cruz, California 95064 \break
  \iCIN
     University of Cincinnati,
     Cincinnati, Ohio 45221 \break
  \iCSU
     Colorado State University,
     Fort Collins, Colorado 80523 \break
  \iCOLO
     University of Colorado,
     Boulder, Colorado 80309 \break
  \iCOL
     Columbia University,
     New York, New York 10027 \break
  \iFER
     INFN Sezione di Ferrara and Universit\`a di Ferrara,
     I-44100 Ferrara, Italy \break
  \iFRA
     INFN  Lab. Nazionali di Frascati,
     I-00044 Frascati, Italy \break
  \iILL
     University of Illinois,
     Urbana, Illinois 61801 \break
  \iLBL
     Lawrence Berkeley Laboratory, University of California,
     Berkeley, California 94720 \break
  \iMIT
     Massachusetts Institute of Technology,
     Cambridge, Massachusetts 02139 \break
  \iMASS
     University of Massachusetts,
     Amherst, Massachusetts 01003 \break
  \iMISS
     University of Mississippi,
     University, Mississippi  38677 \break
  \iNAG
     Nagoya University,
     Chikusa-ku, Nagoya 464 Japan  \break
  \iOREG
     University of Oregon,
     Eugene, Oregon 97403 \break
  \iPAD
     INFN Sezione di Padova and Universit\`a di Padova,
     I-35100 Padova, Italy \break
  \iPERU
     INFN Sezione di Perugia and Universit\`a di Perugia,
     I-06100 Perugia, Italy \break
  \iPISA
     INFN Sezione di Pisa and Universit\`a di Pisa,
     I-56100 Pisa, Italy \break
  \iRUT
     Rutgers University,
     Piscataway, New Jersey 08855 \break
  \iRAL
     Rutherford Appleton Laboratory,
     Chilton, Didcot, Oxon OX11 0QX United Kingdom \break
  \iSOGANG
     Sogang University,
     Seoul, Korea \break
  \iSOONG
     Soongsil University,
     Seoul, Korea 156-743 \break
  \iSLAC
     Stanford Linear Accelerator Center, Stanford University,
     Stanford, California 94309 \break
  \iTENN
     University of Tennessee,
     Knoxville, Tennessee 37996 \break
  \iTOH
     Tohoku University,
     Sendai 980 Japan \break
  \iVAND
     Vanderbilt University,
     Nashville, Tennessee 37235 \break
  \iWASH
     University of Washington,
     Seattle, Washington 98195 \break
  \iWISC
     University of Wisconsin,
     Madison, Wisconsin 53706 \break
  \iYALE
     Yale University,
     New Haven, Connecticut 06511 \break
  \dead
     Deceased \break
  \andgen
     Also at the Universit\`a di Genova \break
  \andper
     Also at the Universit\`a di Perugia \break
\end{center}